\documentclass[12pt]{article}
\usepackage{graphicx}

\usepackage{amsmath,amssymb}
\usepackage{type1cm}
\usepackage{bm}

\newcommand\beq{\begin{equation}}
\newcommand\eeq{\end{equation}}
\newcommand\beqa{\begin{eqnarray}}
\newcommand\eeqa{\end{eqnarray}}

\newcommand{\ds}[1]{#1 \hspace{-0.5em}/}  % For small character
\newcommand\bk{{\bf k}}
\newcommand\bq{{\bf q}}

\newcommand\bp{{\bf p}}

\begin{document}
\title{Magnetic susceptibility of quark matter within Fermi-liquid theory}

\author{T. Tatsumi and K. Sato}
%\email{tatsumi@ruby.scphys.kyoto-u.ac.jp}
\date{}
\maketitle
\begin{center}
\noindent{\it Department of Physics, Kyoto University, Kyoto, 606-8502, Japan}
\end{center}

\begin{abstract}
Possibility of spontaneous magnetization in QCD and  magnetic properties of
 quark 
matter is discussed by evaluating the magnetic 
 susceptibility within Fermi-liquid theory. The screening effects for
 gluons are taken into account to figure out the specific 
properties of the magnetic transition in gauge theories. It is shown that
 the static screening effect in terms of the Debye mass does not 
necessarily work against the magnetic instability; it promotes the
 instability, depending on the coupling constant and the number of flavors.   
\end{abstract}

Many efforts have been devoted to explore the QCD diagram on the density
($\rho_B$)- temperature ($T$) plane, which is closely related to
relativistic heavy-ion collisions, early universe and compact stars
\cite{sch05}.  In particular, it would be interesting to consider the low
temperature case, where quark degrees of freedom are dominant over
gluons and the concept of the Fermi sea is well defined.
Various correlations are expected to be realized there as in
condensed-matter physics; actually there have been many works about color
superconductivity at high-density quark matter \cite{sch05}.

A possibility of spontaneous spin polarization in quark matter
has been suggested by one of the authors \cite{tat00}, using the
one-gluon-exchange (OGE) interaction. The onset mechanism is very simple
but it reflects an interesting aspect of magnetic phase transition in
gauge theories as QED or QCD \cite{tat00,nak03,her66,nie05}. The direct term gives a
null contribution
due to the charge (color) neutrality, so that the leading-order contribution
comes from the exchange term. The Fock exchange interaction then favors the
spin alignment due to the Pauli principle and its infinite-ranginess gives a particular
density dependence of the exchange term.
If it is realized,
it may give a microscopic origin of strong magnetic field observed in
compact stars, especially magnetars \cite{har06}. 

In this work we further explore 
the properties of the magnetic phase
transition in QCD by examining the magnetic properties of quark matter
in a paramagnetic phase, by the use of
Landau Fermi-liquid (FL) theory \cite{bay04,bay76}. We discuss magnetic susceptibility in
detail  to figure out the
critical behavior of the magnetic phase transition in gauge theories 
at finite density. We only consider here the normal phase of quark
matter as a first step for fully understanding the coexistence of
magnetic and superconducting phases.

It is well-known that there appear infrared (IR) singularities in the FL
interactions in the gauge theories. To improve the IR behavior of the
gauge interactions it is necessary to take into account the screening effects for
the gauge field \cite{her66,bay04}. The inclusion of the screening effects is also required
by the argument of the HDL resummation \cite{sch99}. Since the IR singularities
appear for quasi-particles with the co-linear momenta, the soft
gluon should give a dominant contribution. Then the particle-hole polarization function
$\sim O(g^2\mu^2)$ should be the same order of magnitude with the
energy-momentum of soft gluons $\sim O(g\mu)$, where $g$ and $\mu$ are
the QCD coupling constant and quark chemical potential, respectively. 
For the longitudinal gluons we can see the static
screening by the Debye
mass and the IR behavior is surely improved. However, there is no
static screening for the transverse gluons and there is only the dynamic
screening. Thus the IR singularities are still
left for the interactions through the exchange of the transverse gluons.

Applying the uniform but tiny magnetic field on quark matter, one may evaluate 
magnetic susceptibility defined as
\beq
\chi_M\equiv\sum_{f=u,d,s}\chi_M^f=\sum_{f=u,d,s}\left.\frac{\partial\langle M\rangle_f}{\partial B}\right|_{B=0}
\eeq
with the magnetization $\langle M\rangle_f=V^{-1}\langle \mu_q\int d^3x{\bar
q_f}[(-i{\bf r}\times \nabla)_z+\Sigma_z]q_f\rangle$, where we take ${\bf B}//\hat {\bf z}$.  
In the following we consider the isotropic quark matter, so that the
first orbital contribution can be discarded and only spin degrees of
freedom contributes to magnetization.
Since gluons only carries color, we can consider the partial $\chi_M^f$ for
each flavor
\footnote{We, hereafter,  omit the subscript to specify the flavor for simplicity.}
. The inverse of $\chi_M^f$ measures curvature of the free energy
$F(\langle M\rangle_f)$ at the origin, 
$(\chi_M^f)^{-1}=d^2F/d\langle M\rangle_f^2|_{\langle M\rangle_f=0}$, so that the divergence of 
$\chi_M^f$ signals the spontaneous spin polarization. 
The spin
degrees of freedom of a quark with momentum $\bk$ is specified 
by the spin vector $a_\mu (\bk)$
\cite{tat00}, ${\bf a}(\bk)=\zeta[{\hat {\bf z}}+\bk({\hat {\bf
z}}\cdot\bk)/m(E_k+m)], a^0(\bk)=\zeta{\hat {\bf z}}\cdot\bk/m$ with
$E_k=(\bk^2+m^2)^{1/2}$ and $\zeta=\pm 1$. 
We consider the static
susceptibility for a constant magnetic field along $z-$ axis. Each Fermi
sea for $\zeta=\pm 1$ is a little changed in the presence of the magnetic field.
Then magnetization may be given by the difference in the number
densities 
$\delta n_\bk^a=n_{\bk,\zeta=1}-n_{\bk,
\zeta=-1}$ with the distribution function, 
$n_{\bk,\zeta}=[{\rm
exp}\beta(\epsilon_{\bk,\zeta}-\mu-1/2g_D(\bk)\mu_q\zeta B)+1]^{-1}$,
where $\epsilon_{\bk,\zeta}$ is the quasi-particle energy in the
presence of the external magnetic field
\footnote{Generally topology of the Fermi surface is modified due to the
spin-momentum coupling, but one may safely discard it in the
weak-field limit. }
. Thus we find 
\beq
\langle
M\rangle=\frac{\mu_q}{2}N_c\int\frac{d^3k}{(2\pi)^3}g_D(\bk)\delta n_\bk^a
\eeq
where $\mu_q$ denotes the Dirac magnetic
moment, $\mu_q=e_q/2m$, and the gyromagnetic ratio $g_D(\bk)$ is defined
as $g_D(\bk)\equiv 2(a_z-k_z/E_k a_0)=2[1-k_z^2/E_k(E_k+m)]$. For the weak
magnetic field, dynamics among the quasi-particles around the Fermi
surface determine the response to the external magnetic field. Actually 
we find that the number-difference arises as a result of change of the
quasi-particle energy around the Fermi surface,
\beqa
\delta n_\bk^a&=&\sum_\zeta\frac{\partial
n_\bk}{\partial\epsilon_\bk}\left(\epsilon_{\bk,\zeta}-\epsilon_\bk\right)\nonumber\\
&=&\frac{\partial
n_\bk}{\partial\epsilon_\bk}\left[-g_D(\bk)\mu_qB+N_c\sum_{\zeta,\zeta'}\zeta
\int\frac{d^3q}{(2\pi)^3}f_{\bk\zeta,\bq\zeta'}\delta n_{\bq,\zeta'}\right],
\label{FL}
\eeqa 
where $\epsilon_\bk$ denotes the quasi-particle energy in the absence of
the magnetic field.
Here the first term imply the explicit $B$ dependence of the
quasi-particle energy, while the second term is a rearrangement term 
peculiar to FL theory 
and represent the implicit change
of the quasi-particle energy through the change of the distribution
function, 
$\delta n_{\bq,\zeta}\equiv
n_{\bq,\zeta}-n_\bk$. Thus we may see that some non-perturbative effects are
taken into account in Eq.~(\ref{FL}) in a self-consistent manner, even if 
we treat the
quasi-particle interaction $f_{\bk\zeta,\bq\zeta'}$ in a perturbative
way. The quasi-particle interaction  
consists of two parts, the spin independent ($f^s_{\bk,\bq}$) and
independent ($f^a_{\bk,\bq}$) terms;
\beq
f_{\bk\zeta,\bq\zeta'}=f^s_{\bk,\bq}+\zeta\zeta'f^a_{\bk,\bq}.
\eeq
From Eqs.~(2),(3) magnetic susceptibility is written in terms of the FL
interaction,
\beqa
\chi_M=\left(\frac{{\bar g}_D\mu_q}{2}\right)^2\frac{N(T)}{1+N(T)\bar
f^a}
\eeqa
with ${\bar g}_D\equiv \int_{|\bk|=k_F}d\Omega_{\bk}/4\pi g_D(\bk)$.
$\bar f^a$ is an angle-averaged Landau-Migdal parameter, which measures
the strength of the spin dependent interaction,
\beq
{\bar f^a}\equiv \int\frac{d\Omega_\bk}{4\pi}\int\frac{d\Omega_\bq}
{4\pi}\left.f^a_{\bk,\bq}\right|_{|\bk|=|\bq|=k_F}
\eeq
at $T=0$.

$N(T)$ is the effective density of states at the Fermi surface,
\beq
N(T)=-2N_c\int\frac{d^3k}{(2\pi)^3}\frac{\partial
n_\bk}{\partial\epsilon_\bk}, 
\eeq 
and is simply written as 
\beq
N^{-1}(0)=\frac{\pi^2}{N_ck_F^2}v_F
\label{velocity}
\eeq
in the limit of zero temperature . Eq.~(\ref{velocity})
defines the Fermi velocity, which is given by the using the Lorentz
transformation \cite{bay76},
\beq
v_F\equiv \left.\frac{\partial
n_\bk}{\partial\epsilon_\bk}\right|_{|\bk|=k_F}
=\frac{k_F}{\mu}-\frac{N_ck_F^2}{3\pi^2}f_1^s,
\eeq 
 where $f_1^s$ is a spin-averaged Landau-Migdal parameter defined by 
\beq
f_1^s=\left.\frac{3}{4}\sum_{\zeta,\zeta'}\int\frac{d\Omega_{\hat{\bk\bq}}}{4\pi}\cos\theta_{\hat{\bk\bq}} 
f^s_{\bk,\bq}\right|_{|\bk|=|\bq|=k_F},
\eeq
with the relative angle $\theta_{\hat{\bk\bq}}$ of $\bk$ and $\bq$.

Finally the magnetic susceptibility at zero temperature can be written in terms of the
Landau-Migdal parameters,
\beq
\chi_M=\chi_{\rm Pauli}\left[1+\frac{N_ck_F\mu}{\pi^2}\left(-\frac{1}{3}f_1^s+\bar
f^a\right)\right]^{-1},
\eeq
where $\chi_{\rm Pauli}$ is the usual one for the Pauli paramagnetism,
$
\chi_{\rm Pauli}={\bar g}_D^{2}\mu_q^2N_ck_F\mu/4\pi^2.
$

When we consider the color-symmetric forward scattering amplitude of the
two quarks around the Fermi surface by the one gluon
exchange interaction (OGE), the direct term should be vanished due to
the color neutrality of quark matter and the Fock exchange term gives a
leading contribution.  
The color-symmetric OGE interaction of quasi-particles may be written,
\beq
f_{\bk\zeta,\bq\zeta'}
=\frac{1}{N_c^2}\sum_{a,b}f_{\bk\zeta
a ,\bq\zeta' b }
=\frac{m}{E_k}\frac{m}{E_q}M_{\bk\zeta,\bq\zeta'},
\eeq
with the invariant matrix element,
\beq
M_{\bk\zeta,\bq\zeta'}=-g^2\frac{1}{N_c^2}{\rm
tr}\left(\lambda_a/2\lambda_a/2\right)M^{\mu\nu}(k,\zeta; q,\zeta')D_{\mu\nu}(k-q),
\eeq
where $M^{\mu\nu}(k,\zeta; q,\zeta')
=1/(4m^2){\rm
tr}\left[\gamma^\mu(\ds{k}+m)P(a(\bk))\gamma^\nu(\ds{q}+m)P(a(\bq))\right]$;
$P(a)$ is the projection operator to select a state with spin vector
$a^\mu$ \cite{tat00,nak03}.
It has been well known that massless gluons often causes infrared
divergences in the Landau parameters \cite{bay76,tat07}.

Since the one gluon exchange interaction is a long-range force and we
consider the small energy-momentum transfer between quasi-particles, we
must treat the gluon propagator by taking into account the hard-dense-loop (HDL)
resummation \cite{sch99}. Thus we take into account the screening effect
for gluons,
\beq
D_{\mu\nu}(k-q)=P^T_{\mu\nu}D_T(p)+P^L_{\mu\nu}D_L(p)-\xi\frac{p_\mu
p_\nu}{p^4} 
\eeq
with $p=k-q$, where $P^{T(L)}_{\mu\nu}$ is the standard projection operator onto the
transverse (longitudinal) mode \cite{sch99}, which propagator has the form 
$D_{T(L)}(p)=(p^2-\Pi_{T(L)}(p))^{-1}$ in terms of the self-energy $\Pi_{T(L)}(p)$. 
The last term represents the gauge dependence with a parameter
$\xi$. 
First of all we can easily check the gauge independence of the interaction
between 
quasi-particles. Since $p_\mu p_\nu M^{\mu\nu}=0$ for the on-shell
particles, any
physical quantity described by the quasi-particle interaction is
independent of the gauge parameter $\xi$ as should be. 

The self-energies for the transverse and longitudinal gluons are given
as 
\beqa
\Pi_L(p_0,{\bf p})&=&m_D^2\nonumber\\
\Pi_T(p_0,{\bf p})&=&-i\frac{\pi u_F
m_D^2}{4}\frac{p_0}{|\bp|},~~~u_F\equiv \frac{k_F}{E_F}, 
\eeqa
in the limit $p_0/|\bp|\rightarrow 0$, with the Debye screening mass,
$m_D^2\equiv \sum_{f=u,d,s}(1/2\pi^2)g^2\mu_f k_F^f$ \cite{sch99}. 
Thus the longitudinal gluon is
screened to have the Debye mass $m_D$, while the transverse gluon
receives only the dynamic screening and the Fermi-liquid interaction on
the Fermi surface is not screened. 
Then the quasiparticle interaction on the Fermi surface can be written as 
\beqa
\left.f_{\bk\zeta,\bq\zeta'}\right|_{|\bk|=|\bq|=k_F}&=&-C_g\frac{m^2}{E_F^2}
\left[-M^{00}D_L(\bk-\bq)+M^{ii}D_T(\bk-\bq)\right],
\eeqa
with the effective coupling strength, 
$C_g=\frac{N_c^2-1}{2N_c^2}g^2$.

We can see that the both Landau parameters $f_1^s,
{\bar f}^a$ include
the infrared singularities due to the absence of the static screening
for the transverse gluons; $D_T(\bk-\bq)\sim
-1/(\bk-\bq)^2=-1/2k_F^2(1-\cos\theta_{\hat{\bk\bq}})$ in this case, so
that the logarithmic divergences appear in the Landau parameters through
the integral over the relative angle, $\int
d\Omega_{\hat{\bk\bq}}1/(1-\cos\theta_{\hat{\bk\bq}})$.

Finally magnetic susceptibility is given as
a sum of the contributions of the bare interaction and the static 
screening effect. We can see that the logarithmic divergences exactly 
cancel each other to give a finite result for susceptibility.
\begin{equation}
\left(\chi_M/ \chi_{\rm Pauli}\right)^{-1}=1
-\frac{C_gN_c\mu}{12\pi^2E_F^2k_F}\Big[m(2E_F+m)
-\frac{1}{2}(E_F^2+4E_Fm-2m^2)
\kappa\ln\frac{2}{\kappa} \Big],
\label{final}
\end{equation}
with $\kappa=m_D^2/2k_F^2$.
Obviously this expression is reduced to the simple OGE case without
screening in the limit $\kappa\rightarrow 0$; 
one can see that the interaction among massless quarks gives 
a null contribution for the magnetic transition. 
The effect of the static screening for the
longitudinal gluons gives the contribution of $g^4\ln(1/g^2)$. In the
nonrelativistic limit, it recovers the corresponding term in the RPA
calculation of electron gas \cite{her66,bru57}.

\begin{figure}[h]
\begin{center}
\includegraphics[width=0.6\textwidth]{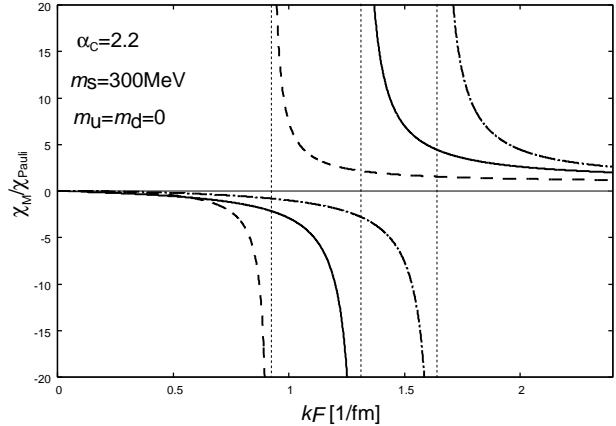}
\caption{Magnetic susceptibility at $T=0$. Screening effects are shown
 in comparison with the simple OGE case: the solid curve shows the result
 with the simple OGE without screening, while the dashed and dash-dotted ones 
 shows the screening effect with $N_f=1$ (only $s$ quark)and $N_f=2+1$
 ($u,d,s$ quarks), respectively.}
\end{center}
\end{figure}

To demonstrate the screening effect, we plot in Fig.~1 the magnetic
susceptibility. We assume a flavor-symmetric quark matter, 
$\rho_u=\rho_d=\rho_s=\rho_B/3$, and take the
QCD coupling constant as $\alpha_c\equiv g^2/4\pi=2.2$ from the MIT bag
model, where the value of $\alpha_c$ is chosen so as to reproduce the
mass difference among different spin states. Since we can see from Eq.~(\ref{final})
that massless quarks, $u$ and $d$ quarks, never lead to magnetic 
phase transition, we
only consider the magnetic susceptibility of massive quarks, $s$ quark
with mass of $300$ MeV in the MIT bag model. Note that all the flavors
always take part in the Debye mass, even if we are only concerned with
the magnetic susceptibility of $s$ quarks.

One can see that the magnetic susceptibility diverges 
around $k_F=1\sim 1.5$fm$^{-1}$ corresponding to order of nuclear
density, where spontaneous magnetization appears. The critical density
 for the 
simple OGE without screening is consistent with the previous one given  
by the energy calculation \cite{tat00}. One may expect that the quasi-particle interaction
is effectively cut off at momentum $|\bk|=m_D$, which reduces the
strength of the Fock exchange interaction, once the screening is taken
into account. However, this is not necessarily the case in QCD.
Compare the screening
effects by changing the number of flavors ($N_f$); if we change $N_f$, the screening
effect exhibits the opposite behavior for $N_f=1$ (only $s$ quark) and
$N_F=2+1$ ($u,d,s$ quarks). 
In the case
of $N_f=1$ the screening effect works against the magnetic phase
transition as in QED \cite{her66}, while it favors in the case of $N_f=2+1$.
This occurs due to the sign change of the logarithmic term in Eq.~(\ref{final}) at
$\kappa=2$ (see Fig.~2): if $\alpha_c N_f>2\pi$, $\kappa>2$ and the screening effect 
acts for the magnetic instability. 
\footnote{It would be interesting to consider the criterion in the large
$N_c$ and $N_f$ limits \cite{ohn07}; since $\alpha_c N_f\sim O(N_f/N_c)$, the
screening effect changes its property, depending on the ratio $N_f/N_c$.}
Hence we may say that the screening
effect does not necessarily work against the magnetic instability in QCD
due to the large coupling constant and flavor degree of freedom.

\begin{figure}[h]
\begin{center}
\includegraphics[width=0.6\textwidth]{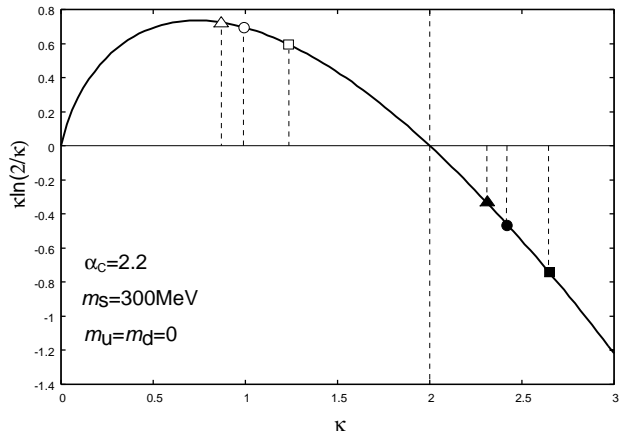}
\caption{Behavior of the logarithmic function is sketched as a function
 of the ratio, $\kappa=m_D^2/2k_F^2$. Empty (full) symbols denote the
 case of $N_f=1 (2+1)$ at $1.0, 1.5, 2.0$fm$^{-1}$ for square, circle,
 triangle, respectively. $\kappa=2$ is the point 
 characterizing the change of the qualitative behavior of susceptibility;
 $\kappa$ for $N_f=1$ and $N_f=2+1$ lie in the different sides.}
\end{center}
\end{figure}

We have discussed the properties of the magnetic phase transition in
QCD, by examining the magnetic susceptibility. The interesting behavior
of the screening effect has been demonstrated, but it comes from the
static screening for the longitudinal gluons. There is no static screening
for the transverse gluons and the dynamic screening never contributes to
the magnetic phase transition at $T=0$ due to the sharpness of the Fermi
surface. However, one may expect that the
dynamic screening gives a finite contribution due to the diffuseness of
the Fermi surface. Actually it is well known that the non-Fermi-liquid
effect gives an anomalous term $\propto T\ln T$ to the specific heat, which is
responsible to the dynamic screening \cite{ger04}. We will consider the
susceptibility in the finite temperature and discuss this issue
in another paper \cite{sat08}. 
 
This work is partially supported by the 
Grant-in-Aid for the 21st Century COE
``Center for the Diversity and Universality in Physics'' 
and the 
Grant-in-Aid for Scientific Research Fund 
of the Ministry of Education, Culture, Sports, Science and Technology of Japan
(13640282, 16540246).

\end{document}